\documentclass[10pt,fleqn]{article}
\usepackage{amssymb}

\newcommand{\name}[1]{\begin{flushleft}
                       \LARGE \bf #1
                       \end{flushleft}\vspace{-3mm}}

\newcommand{\Author}[1]{\begin{flushleft}
                       \it #1 \end{flushleft}}

\newcommand{\Adress}[1]{\begin{flushleft}
                       \it #1 \end{flushleft}}

\newcommand{\be}{\begin{equation}}
\newcommand{\ee}{\end{equation}}
\newcommand{\ba}{\hspace*{-5pt}\begin{array}}
\newcommand{\ea}{\end{array}}
\newcommand{\p}{\partial}
\newcommand{\ds}{\displaystyle}

\newcommand{\pbf}[1]{\mbox{\mathversion{bold}$#1$}}

\begin{document}

\name{$\pbf{P}$, $\pbf{T}$, $\pbf{C}$ properties of the Poincar\'e invariant equations for
massive particles}

\medskip

\noindent{published in {\it Lettere al Nuovo Cimento}, 1973, {\bf 6}, N~4, P. 133--137.}

\Author{Wilhelm I. FUSHCHYCH\par}

\Adress{Institute of Mathematics of the National Academy of
Sciences of Ukraine, \\ 3 Tereshchenkivska  Street, 01601 Kyiv-4,
UKRAINE}

\noindent {\tt URL:
http://www.imath.kiev.ua/\~{}appmath/wif.html\\ E-mail:
symmetry@imath.kiev.ua}

\bigskip

Recently [1] we have shown that for free particles and antiparticles with mass $m>0$ and
arbitrary spin $s>0$, in the framework of the Poincar\'e group $P(1,3)$,
there exist three types of nonequivalent equations. In the present paper we study
the $P$, $T$, $C$ properties of these equations.

It will be convinient to investigate these properties in the canonical representation where
the Hamiltonian is diagonal (as matrix) and other operators (position operator and
spin operator) have adequate physical interpretation. For the transformation to this
representation let us make unitary transformation~[2]
\be
\ba{l}
\ds {\mathcal U}\left(p,s=\frac 12\right) =\exp\left[\frac{\pi}{4}\frac{\Gamma_0{\mathcal H}^{(8)}}{E}
\right]=\frac{1}{\sqrt{2}} \left(1+\frac{\Gamma_0{\mathcal H}^{(8)}}{E}\right),
\vspace{2mm}\\
\ds {\mathcal H}^{(8)}\equiv \Gamma_0 \Gamma_k p_k, \qquad p_4\equiv m, \qquad k=1,2,3,4,
\ea
\ee
over the eight-component equation of the Dirac type
\be
i\frac{\p\Psi^{(8)}(t,\pbf{x})}{\p t}={\mathcal H}^{(8)}\Psi^{(8)}(t,\pbf{x}).
\ee
Equation (2) after the transformation (1) transfers into
\be
i\frac{\p\Phi^{(8)}(t,\pbf{x})}{\p t}={\mathcal H}^{c}\Phi^{(8)}(t,\pbf{x}), \qquad
{\mathcal H}^c =\Gamma_0 E, \qquad \Phi^{(8)}=U\Psi^{(8)}.
\ee
In the canonical representation the generators of the $P(1,3)$ group have the form~[2]
\be
\ba{l}
\ds P_0={\mathcal H}^c=\Gamma_0 E, \qquad P_a=p_a =-i\frac{\p}{\p x_a}, \qquad a=1,2,3,
\vspace{2mm}\\
\ds J_{ab}=M_{ab}+S_{ab}, \qquad M_{ab}=x_ap_b -x_b p_a,
\vspace{2mm}\\
\ds J_{0a}=x_0 p_a -\frac 12[x_a,{\mathcal H}^c]_+-\Gamma_0\frac{S_{ab} p_b +S_{04}m}{E},
\qquad x_0\equiv t,
\ea
\ee
where $S_{ab}$, $S_{04}$ matrices are generators of the $SO_4\sim SU_2\otimes SU_2$ group.
On the solutions $\{\Phi^{(8)}\}$ of eq.(2) thhese matrices have form
\[
S_{kl}=S_{kl}^{(8)}=\frac i4 (\Gamma_k \Gamma_l-\Gamma_l \Gamma_k), \qquad k,l=1,2,3,4.
\]
The representation for the generators $P(1,3)$ in the form (4) differs from the
Foldy--Shirokov [3,~4] representation. In the form (4) it is explicity distinguished the fact
that in the space where a representation of the $P(1,3)$ group is given, also a
representation of $SO_4\sim SU_2\otimes SU_2$ is realized. This follows,
in particular, from the fact $[{\mathcal H}^c, S_{kl}]_-=0$, i.e. it means that the matrices
\[
S_a=\frac 12\left(\frac 12\varepsilon_{abc}S_{bc}+S_{4a}\right), \qquad
T_a=\frac 12\left(\frac 12\varepsilon_{abc}S_{bc}-S_{4a}\right),
\]\renewcommand{\thefootnote}{*}
commute with the Hamiltonian\footnote{In fact, eq.(2) or (3) is invariant with respect to
$SO_6\supset SO_4$ group~[2]. A relativistic equation of motion for particle with spin~$\frac 32$
is invariant also with respect to the $SO_6$ group.}. In other words this means that the
space, where the representation of $P(1,3)$ group is realized, must be characterized (besides
the mass $m$ and the sign of the energy) by pair of indices $s$ and $\tau$
\[
S_a^2\Phi=s(s+1)\Phi, \qquad T_a^2 \Phi=\tau(\tau+1)\Phi,
\qquad s,\tau=\frac 12, 1,\frac 32,\ldots.
\]
we shall denote by $D^\pm(s,0)$ and $D^\pm(0,\tau)$ the irreducible representation of $P(1,3)$
group. For futher understanding it should be noted that the irreducible representations
$D(s,0)$ and $D(s,0)$ of $SO_4$ group are indistinguishable with respect to the matrices $S_{ab}$
from the $SO_3$ algebra.

From the canonical eight-component equation (3) we can obtain the following three types
of nonequivalent four-component equations
\be
i\frac{\p \Phi_a(t,\pbf{x})}{\p t}={\mathcal H}_a\Phi_a(t,\pbf{x}), \qquad a=1,2,3,
\ee
\be
{\mathcal H}_1={\mathcal H}_2=\varepsilon \gamma_0 E, \qquad {\mathcal H}_3=\varepsilon E,
\qquad \varepsilon=\pm 1,
\ee\renewcommand{\thefootnote}{**}
where $\gamma_0$ is the hermitian and diagonal $4\times 4$ matrix\footnote{The fact
that the ${\mathcal H}_1$ and ${\mathcal H}_2$ have identical forms in two eqs.(5) must
not lead into confusion since the equation of motion is defined completly if only we determine
both the Hamiltonian and the representation of $P(1,3)$ group.}.
Under a transformation of the $P(1,3)$ group the four-component wave functions
$\Phi_1$, $\Phi_2$, $\Phi_3$ transform on the representations (for the sake of
brevity we consider only case $\varepsilon=+1$)
\be
D^+(s,0)\oplus D^-(0,\tau), \qquad s=\tau=\frac 12,
\ee
\be
D^+(s,0)\oplus D^-(s,0), \qquad s=\frac 12, \ \tau=0,
\ee
\be
D^+(s,0)\oplus D^+(0,\tau), \qquad s=\tau=\frac 12.
\ee
On the manifolds $\{\Phi_1\}$, $\{\Phi_2\}$, $\{\Phi_3\}$ the generators $P_\mu$,
$J_{\alpha\beta}$ have the forms
\be
\ba{l}
\ds P_0^{(1)}={\mathcal H}_1, \qquad P_a^{(1)}=p_a, \qquad J_{ab}^{(1)}=M_{ab}+S_{ab},
\vspace{2mm}\\
\ds J_{0a}^{(1)}=x_0 p_a -\frac 12[x_a,{\mathcal H}_1]_+-\gamma_0
\frac{S_{ab}p_b +S_{a4}m}{E};
\ea
\ee
\be
\ba{l}
\ds P_0^{(2)}={\mathcal H}_2, \qquad P_a^{(2)}=p_a, \qquad J_{ab}^{(2)}=M_{ab}+S_{ab},
\vspace{2mm}\\
\ds J_{0a}^{(2)}=x_0 p_a -\frac 12[x_a,{\mathcal H}_2]_+-\gamma_0
\frac{S_{ab}p_b +\frac 12\varepsilon_{abc} S_{bc}m}{E};
\ea
\ee
\be
\ba{l}
\ds P_0^{(3)}={\mathcal H}_3=E, \qquad P_a^{(3)}=p_a, \qquad J_{ab}^{(3)}=M_{ab}+S_{ab},
\vspace{2mm}\\
\ds J_{0a}^{(3)}=x_0 p_a -\frac 12[x_a,E]_+-
\frac{S_{ab}p_b +S_{a4}m}{E}\equiv x_0p_a -x_a E+S_{0a}\frac{{\mathcal H}}{E},
\ea
\ee
where
\[
{\mathcal H}=\gamma_0 \gamma_k p_k, \qquad S_{\mu\nu}=\frac i4(\gamma_\mu \gamma_\nu-
\gamma_\nu\gamma_\mu), \qquad \mu=0,1,2,3,4.
\]

It should be noted that only in the last representation (12) the Hamiltonian ${\mathcal H}_3=E$
is the positive-definite operator. If we add to the algebra (12) an operator of the change
$Q=\gamma_0$, then such algebra (in the quantum mechanics framework) has the same
properties as the corresponding Poincar\'e algebra, obtained by the procedure of the
Dirac equation quantization.

It is well known [3] that there exist two nonequivalent definitions of the space-reflection
operator $P$:
\be
P^{(1)}\Phi(t,\pbf{x},m)=r_1\Phi(t,-\pbf{x}, m), \qquad \left(P^{(1)}\right)^2\sim 1,
\ee
\be
P^{(2)}\Phi(t,\pbf{x},m)=r_2\Phi^*(t,-\pbf{x}, m), \qquad \left(P^{(2)}\right)^2\sim 1,
\ee
\be
[P^{(1)},P_0]_-=0=[P^{(1)},J_{ab}]_-, \qquad  [P^{(1)},P_a]_+=0=[P^{(1)},J_{0a}]_+,
\ee
\be
[P^{(2)},P_0]_+=0=[P^{(2)},J_{ab}]_+, \qquad  [P^{(2)},P_a]_-=0=[P^{(2)},J_{0a}]_-.
\ee
Also there exist two nonequivalent definitions of the time-reflection $T$:
\be
T^{(1)}\Phi(t,\pbf{x},m)=t_1\Phi(-t,\pbf{x}, m), \qquad \left(T^{(1)}\right)^2\sim 1,
\ee
\be
T^{(2)}\Phi(t,\pbf{x},m)=t_2\Phi^*(-t,\pbf{x}, m), \qquad \left(T^{(2)}\right)^2\sim 1,
\ee
\be
[T^{(1)},P_0]_+=0=[T^{(1)},J_{0a}]_+, \qquad  [T^{(1)},P_a]_-=0=[T^{(1)},J_{ab}]_-,
\ee
\be
[T^{(2)},P_0]_-=0=[T^{(2)},J_{0a}]_-, \qquad  [T^{(2)},P_a]_+=0=[T^{(2)},J_{ab}]_+.
\ee

Besides these conditions usually imposed on the discrete operators $P$ and $T$ we shall
require also the subsidiary conditions
\be
[\hat X_a,P^{(1)}]_+=0=[P^{(2)},\hat X_a]_+,
\ee
\be
[T^{(1)}, \hat X_a]_-=0=[T^{(2)},\hat X_a]_-
\ee
to be satisfied where $\hat X_a$ is a position operator. The conditions (21) and (22)
guarantee that quantities $r_1$, $r_2$, $t_1$, $t_2$ are the matrices which do not depend
on the momentum. If the conditions (21), (22) are not imposed, then the operators $P$ and $T$
may be nonlocal (in this case the quantities depend on the momentum).

In addition to the discrete operators $P$ and $T$ we shall introduce some more discrete
operators:
\be
M\Phi(t,\pbf{x},m)=r_m\Phi(t,\pbf{x}, -m), \qquad M^2\sim 1,
\ee
\be
M_t\Phi(t,\pbf{x},m)=m_t\Phi(-t,\pbf{x}, -m), \qquad M^2_t\sim 1,
\ee
\be
M_x\Phi(t,\pbf{x},m)=m_x\Phi(t,-\pbf{x}, -m), \qquad M^2_x\sim 1,
\ee
\be
[M,P_\mu]_-=0 =[M,J_{\mu\nu}]_-, \qquad \mu,\nu=0,1,2,3,
\ee
\be
[M_t,P_0]_+=0 =[M_t,J_{0a}]_+, \qquad  [M_t,P_a]_-=0 =[M_t,J_{ab}]_-,
\ee
\be
[M_x,P_0]_-=0 =[M_x,J_{ab}]_-, \qquad  [M_x,P_a]_+=0 =[M_x,J_{0a}]_+,
\ee
where $r_m$, $m_t$, $m_x$ are the $4\times 4$ matrices.

There is no need to define specially the operator of the charge conjugation $C$ since
it is equal to the operator $T^{(1)}\cdot T^{(2)}$ (or $P^{(1)}\cdot P^{(2)}$).

If we use the explicit forms (10)--(28) for the generators $P_\mu$ and $J_{\alpha\beta}$
and carrying out the analysis of the conditions (13)--(28) we come to the following results:

\vspace{-2mm}

\begin{enumerate}\renewcommand{\thefootnote}{*}

\item[1)] Equation (5) for the function $\Phi_1$ (taking into consideration the
representa\-tion~(10)) is $C$, $M_x$, $M_t$, $P^{(1)}T^{(2)}$ invariant, but $P^{(1)}$, $P^{(2)}$,
$T^{(2)}$, $M$ noninvariant;

\vspace{-2mm}

\item[2)] Equation (5) for the function $\Phi_2$ (taking into consideration the
representa\-tion~(11)) is $P^{(2)}$, $T^{(1)}$, $M_x$, $P^{(1)}T^{(2)}$ invariant,
but $P^{(1)}$, $T^{(2)}$, $C$, $M$, $M_t$ noninvariant\footnote{In the coupling scheme,
brought in ref. [1], the correction $D^+(s,0)\; {\mathop{\leftrightarrow }\limits^C}\; D^-(0,s)$
should be done.};

\vspace{-2mm}

\item[3)] Equation (5) for the function $\Phi_3$ (taking into consideration the
representa\-tion~(12)) is $P^{(1)}$, $T^{(2)}$, $M$, $M_x$, $P^{(1)}T^{(2)}$ invariant,
but $T^{(1)}$, $C$, $P^{(2)}$, $M_t$ noninvariant.

\vspace{-2mm}

\end{enumerate}

These assertions may be proved also starting from eight-component equation~(2)
(or~(3)) in which constraints have been imposed on the wave function~[1]. To establish
this it is necessary to analyse the commutation relations between the discrete
opera\-tors and the projections $P_1^\pm$, $P_2^\pm$, $P_3^\pm$.

\smallskip

\noindent
{\bf Note 1.} It can be easily checked that
\be
P^{(1)}S_a=T_a P^{(1)}, \qquad MS_a=T_aM, \qquad T^{(1)}S_a=S_a T^{(1)}.
\ee
The transformation connecting the cannonical representations (10)--(12) and the
Fol\-dy--Shirokov representation has the form
\be
U_1=\frac{m+E+\gamma_4 \gamma_a p_a}{\{2E(E+m)\}^{1/2}}.
\ee

\noindent
{\bf Note 2.} If we put $m=0$ in the reducible representation (4), then it reduces into the
following direct sum of the irreducible representation of the $P(1,3)$ algebra
\be
\ba{l}
\ds D^+\left(\frac 12,0\right)\oplus D^-\left(0,\frac 12\right)\oplus
D^-\left(\frac 12,0\right)\oplus D^+\left(0,\frac 12\right) \to
\vspace{2mm}\\
\ds \qquad \to D^+\left(\frac 12,0\right)\oplus
D^+\left(-\frac 12,0\right)\oplus
D^-\left(0, \frac 12\right)\oplus D^-\left(0,-\frac 12\right)\oplus
\vspace{2mm}\\
\ds \qquad \qquad \oplus
D^-\left(\frac 12,0\right)\oplus D^-\left(-\frac 12,0\right)\oplus D^+\left(0,\frac 12\right)\oplus
D^+\left(0,-\frac 12\right),
\ea
\ee
where members $\frac 12$ and $-\frac12$ are the eigenvalues of the operators $S_ap_a/E$ and
$T_a p_a/E$. These operators commute with the generators $P_\mu$, $J_{\alpha\beta}$
when $m=0$. From (31) follows that there exist 28 types of mathematical nonequivalent
two-component equations for massless particles.

\smallskip

\noindent
{\bf Note 3.} In order that Poincar\'e-invariant equation $m\not=0$ was totally $P$, $T$, $C$
invariant it is necessary and sufficient that the wave function was transformed on the
following direct sum of representation of $P(1,3)$
\be
D^+(s,\tau)\oplus D^-(s,\tau)\oplus D^+(\tau,s)\oplus D^-(\tau,s), \qquad \mbox{if} \quad \tau\not= s,
\ee
\be
D^+(s,\tau)\oplus D^-(s,\tau), \qquad \mbox{if} \quad \tau= s.
\ee
The representation $D^+(s,\tau)$ is in general reducible with respect to the $P(1,3)$ algebra,
therefore the wave function describes a multiplet of particles with variable-spin, but
fixed mass. The spin of the multiplet can take the values from $(s-\tau)$ to $(s+\tau)$.
The equations of motion describing a physical system with variable-mass and
variable-spin were considered in ref.~[5].

\medskip

\begin{enumerate}

\footnotesize

\item Fushchych W.I., {\it Lett. Nuovo Cimento}, 1972, {\bf 4}, 344, {\tt quant-ph/0206104}.

\item Fushchych W.I., {\it Theor. Math. Phys.}, 1971, {\bf 7}, 3; Preprint ITF-70-32, Kyiv, 1970.

\item Foldy L.L., {\it Phys. Rev.}, 1956, {\bf 102}, 568.

\item Shirokov Yu.M., {\it Zurn. Eksp. Teor. Fiz.}, 1957, {\bf 33}, 1196.

\item Fushchych W.I., {\it Theor. Math. Phys.}, 1970, {\bf 4}, 360;\\ Preprint ITF-70-40, Kyiv, 1970, {\tt quant-ph/0206079};\\
Fushchych W.I., Krivsky I.Yu., {\it Nucl. Phys. B}, 1969, {\bf
14}, 573, {\tt quant-ph/0206047}.

\end{enumerate}
\end{document}